\documentclass[
11pt,%
tightenlines,%
twoside,%
onecolumn,%
nofloats,%
nobibnotes,%
nofootinbib,%
superscriptaddress,%
noshowpacs,%
centertags]%
{revtex4}
\usepackage{ljm}
\usepackage{subfigure}

\begin{document}

\titlerunning{Shock Waves in Euler Flows of Gases} % for running heads

\authorrunning{V. Lychagin, M. Roop} % for running heads

\title{Shock Waves in Euler Flows of Gases}
% Splitting into lines is performed by the command \\
% The title is written in accordance with the rules of capitalization.

\author{\firstname{V.~V.}~\surname{Lychagin}}
\email[E-mail: ]{valentin.lychagin@uit.no}
\affiliation{V.A. Trapeznikov Institute of Control Sciences, Russian Academy of Sciences, 65 Profsoyuznaya Str., 117997 Moscow, Russia}

\author{\firstname{M.~D.}~\surname{Roop}}
\email[E-mail: ]{mihail_roop@mail.ru}
\affiliation{V.A. Trapeznikov Institute of Control Sciences, Russian Academy of Sciences, 65 Profsoyuznaya Str., 117997 Moscow, Russia}
\affiliation{Faculty of Physics, Lomonosov Moscow State University, Leninskie Gory, 119991 Moscow, Russia}

%\noaffiliation % If the author does not specify a place of work.

\firstcollaboration{(Submitted by A.~A.~Editor-name)} % Add if you know submitter.
%\lastcollaboration{ }

\received{\today} % The date of receipt to the editor, i.e. December 06, 2017

\begin{abstract} % You shouldn't use formulas and citations in the abstract.
Non-stationary Euler flows of gases are studied. The system of differential equations describing such flows can be represented by means of 2-forms on zero-jet space and we get some exact solutions by means of such a representation. Solutions obtained are multivalued and we provide a method of finding caustics, as well as wave front displacement. The method can be applied to any model of thermodynamic state as well as to any thermodynamic process. We illustrate the method on adiabatic ideal gas flows.
\end{abstract}

 % Enter 2010 Mathematics Subject Classification.

 % Include keywords separeted by comma.

\maketitle

% Text of article starts here.

\section{Introduction}
Critical phenomena in non-stationary gas dynamics, such as shock waves and blow up effects have always been of both theoretical and practical interest. The main difficulty in theoretical investigation of such problems is that corresponding solutions are not smooth to apply numerical methods applicable mostly in the case when solutions are enough smooth functions. The analysis of such effects is provided in, for example, \cite{KorpOvch} for numerous examples of equations of mathematical physics mainly by using methods of functional analysis. Another approach based on geometrical methods \cite{LychSing,KLR,KrVin} is well developed for non-stationary filtration problems, for example, in \cite{AKL-dan, AKL-ifac}, where the system of one-dimensional filtration equations is reduced to Monge-Amp\`{e}re equation that can be solved by means of linearizing Legendre transformation, and in \cite{AKL-gsa}, where a multivalued solution was obtained for rotation-invariant flows.

In this paper, we consider a system of hyperbolic quasilinear equations of the first order, a particular case of Jacobi equations \cite{KLR}, theoretical investigations of global solvability for which are studied in \cite{Tun1,Tun2,Tun3}. Namely, our equations are generalization of those in \cite{Tun3}, where polytropic Euler flows are studied. Comparing with \cite{Tun3}, we do not assume any concrete model of the medium and any concrete process this medium is involved in and provide a method of finding multivalued solutions and constructing discontinuous ones applicable for any thermodynamic model. Our methods are based on representation of Jacobi type systems (see, for example, \cite{KLR}) by means of differential 1-forms on zero-jet space. One of advantages of such a consideration is the reduction of the order of jet-space. We deal with geometrical constructions on zero-jet space instead of one-jet space, where equations in question have natural representation. This idea goes back to \cite{Lych} and has found applications also in incompressible hydrodynamics \cite{RoulRub1,RoulRub2}.

One-dimensional flows of gases are described by the following system of differential equations:
\begin{equation}
\label{Euler}
\begin{cases}
\rho_{t}+(\rho u)_{x}=0,
   \\
  \rho(u_{t}+uu_{x})=-p_{x},
 \end{cases}
\end{equation}
where $u=u(t,x)$ is the velocity of the gas, $\rho=\rho(t,x)$ is the density, $p=p(t,x)$ is the pressure. The first equation is the conservation of mass, and the second one is Euler equation. System (\ref{Euler}) becomes complete once it is extended by equations of state of the medium.

\section{Thermodynamics}
Here, we recall the geometric description of thermodynamic states and processes (see also \cite{LY,LRljm} and references therein). As we will see below, geometrical constructions described here significantly influence system (\ref{Euler}).

Let $(\mathbb{R}^{5},\theta)$ be a contact space with coordinates $(s,e,v,p,T)$ standing for the entropy, the energy, the specific volume $v=\rho^{-1}$, the pressure and the temperature respectively. The contact form $\theta$ is given by
\begin{equation*}
\theta=ds-T^{-1}de-pT^{-1}dv.
\end{equation*}
A thermodynamic state can now be defined as a Legendrian manifold $L\subset(\mathbb{R}^{5},\theta)$, such that $\theta|_{L}=0$. This means that the first law of thermodynamics holds on $L$. By choosing $(e,v)$ as coordinates on $L$, one gets
\begin{equation*}
L=\left\{s=\sigma(e,v),\,T=\frac{1}{\sigma_{e}},\,p=\frac{\sigma_{v}}{\sigma_{e}}\right\},
\end{equation*}
where $\sigma(e,v)$ is a known function. But this function cannot be derived from experiments since there are no ways to measure entropy. To overcome that, we introduce a projection $\pi\colon\mathbb{R}^{5}\to\mathbb{R}^{4}$, where $\pi(s,e,v,p,T)=(e,v,p,T)$ and $\mathbb{R}^{4}(e,v,p,T)$ is equipped with the symplectic form $\Omega=-d\theta$ equal to
\begin{equation*}
\Omega=d(T^{-1})\wedge de+d(pT^{-1})\wedge dv.
\end{equation*}
Then, a thermodynamic state is an immersed Lagrangian manifold $\widehat L=\pi(L)\subset(\mathbb{R}^{4},\Omega)$ that is given by two functions
\begin{equation*}
\widehat L=\left\{f(e,v,p,T)=0,\, g(e,v,p,T)=0\right\},
\end{equation*}
such that $[f,g]=0$ on $\widehat L$, where $[f,g]$ is the Poisson bracket:
\begin{equation*}
[f,g]\Omega\wedge\Omega=df\wedge dg\wedge\Omega.
\end{equation*}
\begin{theorem}
The Lagrangian manifold $\widehat L$ is given by the Massieu-Planck potential $\phi(v,T)$:
\begin{equation}
\label{pe}
p=RT\phi_{v},\quad e=RT^{2}\phi_{T},
\end{equation}
where $R$ is the universal gas constant.

The specific entropy $s$ is of the form
\begin{equation}
\label{ent}
s=R(\phi+T\phi_{T}).
\end{equation}
\end{theorem}

The symplectic space $(\mathbb{R}^{4},\Omega)$ is also equipped with the pseudo-Riemannian structure of signature (2,2) \cite{LY}:
\begin{equation*}
\kappa=d(T^{-1})\cdot de+d(pT^{-1})\cdot dv.
\end{equation*}
The domains on $L$ where its restriction $\kappa|_{\widehat L}$ to the manifold $\widehat L$ is negative are called \textit{applicable states} or \textit{phases}.
\begin{theorem}
The differential quadratic form $\kappa|_{\widehat L}$ is given by the Massieu-Planck potential $\phi(v,T)$:
\begin{equation*}
R^{-1}\kappa|_{\widehat L} =-\left( \phi _{TT}+2T^{-1}\phi _{T}\right) dT\cdot dT+\phi_{vv}dv\cdot dv.
\end{equation*}
\end{theorem}
From the above theorem, (\ref{pe}) and using $v=\rho^{-1}$, we get that applicable domains on $\widehat L$ are given by inequalities
\begin{equation*}
p_{\rho}>0,\quad e_{T}>0.
\end{equation*}
The second inequality holds for a considerable number of thermodynamic models at any point on $L$, while the first one does not. Inequality $p_{\rho}>0$ is exactly what is responsible for phase transitions of the first order (see also \cite{LY,LRljm}).

By a thermodynamic process we mean a contact transformation $\Phi\colon\mathbb{R}^{5}\to\mathbb{R}^{5}$ preserving the Legendrian manifold $L$. Such transformations are generated by contact vector fields tangent to $L$. Let $l\subset L$ be an integral curve of the contact vector field. From now and on, we will call such curve a \textit{thermodynamic process}.
\section{Euler Equations}
\subsection{Hyperbolicity}
Let us assume that thermodynamic state of the gas is given by a Legendrian manifold $L$. And consider the system consisting of (\ref{Euler}), (\ref{pe}) and (\ref{ent}) for a given $\phi(v,T)$. This system becomes complete once we assume that the gas $L$ is involved in some given process $l\subset L$. Indeed, let $\rho$ be a coordinate on $l$. Then, all the thermodynamic variables are known functions of $\rho$, in particular, $p=p(\rho)$. Hence, we will get (\ref{Euler}) as
\begin{equation}
\label{sys}
\begin{cases}
 \rho_{t}+(\rho u)_{x}=0,
   \\
\displaystyle u_{t}+uu_{x}+\frac{p^{\prime}(\rho)}{\rho}\rho_{x}=0,
 \end{cases}
\end{equation}
Let $E=J^{0}(\mathbb{R}^{2})$ with coordinates $(t,x,u,\rho)$ be the space of zero-jets of functions on $M=\mathbb{R}^{2}(t,x)$. Let us associate the following two 2-forms $\omega_{1}, \omega_{2}\in\Lambda^{2}(E)$ with system (\ref{sys}):
\begin{eqnarray*}
\omega_{1}&=&\rho dt\wedge du+udt\wedge d\rho-dx\wedge d\rho,\\
\omega_{2}&=&udt\wedge du+\frac{p^{\prime}(\rho)}{\rho}dt\wedge d\rho-dx\wedge du.
\end{eqnarray*}
Any form $\omega\in\Lambda^{2}(E)$ defines an operator
\begin{equation*}
\Delta_{\omega}\colon C^{\infty}(M)\to\Lambda^{2}(M),\quad \Delta_{\omega}(f)=\omega|_{\Gamma^{0}(f)},
\end{equation*}
where $\Gamma^{0}(f)\subset E$ is a graph of the vector function $f$. We can rewrite system (\ref{sys}) as
\begin{equation*}
\Delta_{\omega_{1}}(f)=0,\quad\Delta_{\omega_{2}}(f)=0,
\end{equation*}
where $f=(u(t,x),\rho(t,x))$. A two-dimensional manifold $N\subset E$ is said to be a \textit{multivalued solution} of (\ref{sys}) if $\omega_{1}|_{N}=\omega_{2}|_{N}=0$.

Let $q=dt\wedge dx\wedge du\wedge d\rho$ be a volume form on $E$. Let us now introduce a bilinear operator
\begin{equation*}
P\colon\Lambda^{2}(E)\times\Lambda^{2}(E)\to C^{\infty}(E)
\end{equation*}
by the following relation
\begin{equation*}
\alpha_{1}\wedge\alpha_{2}=P(\alpha_{1},\alpha_{2})q,\quad\alpha_{i}\in\Lambda^{2}(E).
\end{equation*}
Introduce the notation $P_{\omega}=\|P(\omega_{i},\omega_{j})\|$, $i,j=1,2$. Then, system (\ref{sys}) is said to be \textit{hyperbolic} if $\det(P_{\omega})<0$, \textit{elliptic} if $\det(P_{\omega})>0$ and \textit{parabolic} if $\det(P_{\omega})=0$. Straightforward computations show that
\begin{equation*}
P_{\omega}=
\begin{pmatrix}
2\rho & 0\\
0 & -2\rho^{-1}p^{\prime}(\rho)
\end{pmatrix},
\end{equation*}
and we conclude that condition for (\ref{sys}) to be hyperbolic coincides with the applicability condition of the thermodynamic model, or, equivalently, negativity of the form $\kappa|_{\widehat L}$.
\begin{theorem}
System (\ref{sys}) is hyperbolic if and only if the thermodynamic process curve lies in an applicable domain on $\widehat L$.
\end{theorem}
\subsection{Characteristic distributions}
Note that differential 2-forms
\begin{eqnarray}
\label{forms}
\widehat{\omega_{1}}&=&a_{11}\omega_{1}+a_{12}\omega_{2},\\
\widehat{\omega_{2}}&=&a_{21}\omega_{1}+a_{22}\omega_{2},
\end{eqnarray}
where $a_{11}a_{22}-a_{12}a_{21}\ne 0$ can define system (\ref{sys}) as well. From now and on, we will assume that system (\ref{sys}) is hyperbolic. In this case (see \cite{KLR}) one can choose other forms (\ref{forms}) which we will continue denoting by $\omega_{1}$ and $\omega_{2}$, such that
\begin{equation}
\label{eff}
\omega_{1}\wedge\omega_{2}=0,\quad \omega_{1}\wedge\omega_{1}=-\omega_{2}\wedge\omega_{2}.
\end{equation}
\begin{theorem}
Let (\ref{sys}) be a system of hyperbolic type. Then, it is defined by 2-forms
\begin{eqnarray*}
\omega_{1}&=&A(\rho)(\rho dt\wedge du+udt\wedge d\rho-dx\wedge d\rho),\\
\omega_{2}&=&udt\wedge du+\rho A^{2}(\rho)dt\wedge d\rho-dx\wedge du,
\end{eqnarray*}
where $A(\rho)=\rho^{-1}\sqrt{p^{\prime}(\rho)}$ and 2-forms $\omega_{1}$, $\omega_{2}$ satisfy relations (\ref{eff}).
\end{theorem}
The form $\omega_{2}$ is a closed non-degenerate 2-form which can serve as a symplectic structure on $E$. Then, one can consider the linear operator $A_{\omega}\colon D(E)\to D(E)$ defined by the following way:
\begin{equation*}
X\rfloor\omega_{2}=A_{\omega}(X)\rfloor\omega_{1}.
\end{equation*}
By choosing $\langle\partial_{t},\partial_{x},\partial_{\rho},\partial_{u}\rangle$ as a basis in the module $D(E)$ of vector fields on $E$, we get that the matrix $W$ of the operator $A_{\omega}$ has the following form:
\begin{equation*}
W=\frac{1}{\rho A(\rho)}\begin{pmatrix}
u & -1 & 0 & 0\\
u^{2}-\rho^{2}A^{2}(\rho) & -u & 0 & 0\\
0 &0 & 0 & \rho A^{2}(\rho)\\
0 &0 & \rho & 0
\end{pmatrix},
\end{equation*}
and it is easy to check that $A_{\omega}^{2}=\mathrm{id}$. Eigenspaces $\mathcal{C}_{+}=\langle X_{+},Y_{+}\rangle$ and $\mathcal{C}_{-}=\langle X_{-},Y_{-}\rangle$ of the operator $A_{\omega}$, called \textit{characteristic distributions}, are generated by vector fields
\begin{eqnarray*}
X_{\pm}&=&\pm A(\rho)\partial_{u}+\partial_{\rho},\\
Y_{\pm}&=&(\mp \rho A(\rho)+u)^{-1}\partial_{t}+\partial_{x}.
\end{eqnarray*}
\begin{theorem}
Distributions $\mathcal{C}_{+}$ and $\mathcal{C}_{-}$ are integrable if and only if
\begin{equation}
\label{intr}
p(\rho)=c_{0}\rho^{3}+c_{1},
\end{equation}
where $c_{0}$ and $c_{1}$ are constants.
\end{theorem}
\subsection{Solutions}
Here, we construct multivalued solutions for any gas and any type of process this gas is involved in. Let us look for a 2-dimensional submanifold $N\subset E$, i.e. a multivalued solution, in the form $N\subset N_{1}$, where $N_{1}$ is a 3-dimensional submanifold in $E$. Suppose that $N_{1}$ is given as
\begin{equation*}
N_{1}=\left\{F(t,x,\rho,u)=0\right\}.
\end{equation*}
Let $V_{+}$ and $V_{-}$ be two vector fields from distributions $\mathcal{C}_{+}$ and $\mathcal{C}_{-}$ respectively tangent to $N_{1}$. They have the form
\begin{equation*}
\begin{split}
V_{\pm}&=(F_{u}A(\rho)\pm F_{\rho})\partial_{t}+((uF_{u}-\rho F_{\rho})A(\rho)\pm(uF_{\rho}-\rho A^{2}(\rho)F_{u}))\partial_{x}-{}\\&-(F_{t}+uF_{x}\mp\rho A(\rho)F_{x})\partial_{u}+(\rho A(\rho)F_{x}\mp(F_{t}+uF_{x}))\partial_{\rho}.
\end{split}
\end{equation*}
We need to choose the function $F(t,x,\rho,u)$ in such a way that the distribution $V=\langle V_{-},V_{+}\rangle$ is integrable. This condition leads us to the second order PDE for the function $F(t,x,\rho,u)$, which in case of $x$-independence, i.e. $F(t,x,\rho,u)=f(t,u,\rho)$ has the form
\begin{equation*}
-2\rho A^{2}f_{u}f_{t}f_{ut}-\rho f_{\rho}^{2}f_{tt}-\rho f_{t}^{2}f_{\rho\rho}+\rho A^{2}\left(f_{u}^{2}f_{tt}+f_{t}^{2}f_{uu}\right)+2\rho f_{\rho}f_{t}f_{\rho t}-2f_{\rho}f_{t}^{2}=0,
\end{equation*}
and one of its solutions is
\begin{equation}
\label{usol}
f(u,\rho,t)=\alpha_{0}+\alpha_{1}\rho+\alpha_{2}\rho t-u(\rho+\alpha_{3}),
\end{equation}
where $\alpha_{j}$ are constants.

Let us choose $(t,x,\rho)$ as coordinates on $N_{1}$. Then, restrictions $Z_{\pm}$ of vector fields $V_{\pm}$ to $N_{1}$ will take the form
\begin{equation*}
\begin{split}
Z_{\pm}&=\frac{A(\rho+\alpha_{3})^{2}\mp\alpha_{3}(\alpha_{1}+t\alpha_{2})\pm\alpha_{0}}{(\rho+\alpha_{3})^{2}}\partial_{t}\pm\frac{\alpha_{2}\rho}{\rho+\alpha_{3}}\partial_{\rho}+{}\\&+\frac{\left(A(\rho+\alpha_{3})^{2}\mp\alpha_{3}(\alpha_{1}+t\alpha_{2})\pm\alpha_{0}\right)\left(\mp A\rho(\rho+\alpha_{3})+\rho(t\alpha_{2}+\alpha_{1})+\alpha_{0}\right)}{(\rho+\alpha_{3})^{3}}\partial_{x}.
\end{split}
\end{equation*}

Since the distribution $Z=\langle Z_{+},Z_{-}\rangle$ is integrable, one can easily find its integral and solution for $\rho(t,x)$ is given implicitly by the relation
\begin{equation}
\label{rhosol}
\begin{split}
&\int\frac{A^{2}(\rho+\alpha_{3})}{\alpha_{2}}d\rho+\frac{(\alpha_{1}\alpha_{3}-\alpha_{0})^{2}}{2\alpha_{2}(\rho+\alpha_{3})^{2}}+{}\\&+\frac{2x(\rho+\alpha_{3})^{2}-2t\left(\rho^{2}\alpha_{1}+\alpha_{3}(\alpha_{0}+2\rho\alpha_{1})\right)-\rho\alpha_{2}(\rho+2\alpha_{3})t^{2}}{2(\rho+\alpha_{3})^{2}}=0
\end{split}
\end{equation}
for any function $A(\rho)$. Solution for the velocity can be obtained by means of (\ref{usol}):
\begin{equation}
\label{usol1}
u=\frac{\alpha_{2}\rho t+\alpha_{1}\rho+\alpha_{0}}{\rho+\alpha_{3}}.
\end{equation}

It is worth to say that solution $N$ defined by (\ref{rhosol}) and (\ref{usol1}) is, in general, multivalued. Singularities of projection of $N$ to $\mathbb{R}^{2}(t,x)$ are points where $x_{\rho}=0$, such curve is called \textit{caustic}. Solving equation (\ref{rhosol}) with respect to $x$ and choosing $\rho$ as a coordinate on caustic, we get the following equations for caustic:
\begin{eqnarray}
\label{caus}
x(\rho)&=&-\int\frac{A^{2}(\rho+\alpha_{3})}{\alpha_{2}}d\rho+\frac{\rho(\rho+2\alpha_{3})(\rho+\alpha_{3})^{2}A^{2}-\alpha_{3}^{2}\alpha_{1}^{2}+\alpha_{0}^{2}\pm2\alpha_{0}(\rho+\alpha_{3})^{2}A}{2\alpha_{3}^{2}\alpha_{2}},\\
t(\rho)&=&\frac{\pm A(\rho+\alpha_{3})^{2}-\alpha_{1}\alpha_{3}+\alpha_{0}}{\alpha_{3}\alpha_{2}}.
\end{eqnarray}

To get a discontinuous solution from the multivalued one, one needs to obtain a conservation law. Let us rewrite the continuity equation using (\ref{usol1}):
\begin{equation*}
\rho_{t}+\left(\rho\frac{\alpha_{2}\rho t+\alpha_{1}\rho+\alpha_{0}}{\rho+\alpha_{3}}\right)_{x}=0,
\end{equation*}
from what it follows that we have the conservation law in the form
\begin{equation*}
\Theta=\rho dx-\rho\frac{\alpha_{2}\rho t+\alpha_{1}\rho+\alpha_{0}}{\rho+\alpha_{3}}dt.
\end{equation*}
On our solution the form $\Theta$ is closed and therefore is locally exact. Let us find explicitly its potential. To this end, let us choose $(\rho,t)$ as coordinates on the multivalued solution $N$ given by (\ref{rhosol}) and (\ref{usol1}). In these coordinates, by means of (\ref{rhosol}) we have $x=g(\rho,t)$ and also from (\ref{usol1}) $u=U(\rho,t)$, where
\begin{eqnarray*}
g(\rho,t)&=&-\int\frac{A^{2}(\rho+\alpha_{3})}{\alpha_{2}}d\rho+\frac{\alpha_{2}^{2}t^{2}\rho(\rho+2\alpha_{3})+2t\alpha_{2}\left(\alpha_{3}(\alpha_{0}+2\alpha_{1}\rho)+\alpha_{1}\rho^{2}\right)-(\alpha_{0}-\alpha_{1}\alpha_{3})^{2}}{2\alpha_{2}(\rho+\alpha_{3})^{2}},\\
U(\rho,t)&=&\frac{\alpha_{2}\rho t+\alpha_{1}\rho+\alpha_{0}}{\rho+\alpha_{3}}.
\end{eqnarray*}

The restriction of the form $\Theta$ to $N$ is
\begin{equation*}
\Theta|_{N}=\rho g_{\rho}d\rho+\rho(g_{t}-U(\rho,t))dt.
\end{equation*}
Let $H(\rho,t)$ be a potential of the form $\Theta|_{N}$, i.e. $\Theta|_{N}=H_{\rho}d\rho+H_{t}dt$. Solving an overdetermined system for $H(\rho,t)$
\begin{equation*}
H_{\rho}=\rho g_{\rho},\quad H_{t}=\rho(g_{t}-U(\rho,t)),
\end{equation*}
we get
\begin{equation*}
H(\rho,t)=\frac{(\alpha_{2}\rho t-\alpha_{1}\alpha_{3}+\alpha_{0})\left(\alpha_{1}\alpha_{3}^{2}+\alpha_{3}((t\alpha_{2}+2\alpha_{1})\rho-\alpha_{0})-2\rho\alpha_{0}\right)}{2\alpha_{2}(\rho+\alpha_{3})^{2}}-\int\frac{\rho(\rho+\alpha_{3})A^{2}}{\alpha_{2}}d\rho.
\end{equation*}

Thus to get points where solution has a discontinuity, i.e. a front of shock wave, one needs to resolve the following system
\begin{equation*}
H(\rho_{1},t)=H(\rho_{2},t),\quad g(\rho_{1},t)=g(\rho_{2},t).
\end{equation*}
for a given moment $t$.
\subsection{Ideal gas}
Here, we illustrate described above method of finding caustic and shock wave front for adiabatic flows of ideal gases.

The Legendrian manifold for ideal gases is given by the following state equations:
\begin{equation*}
p=\frac{RT}{v},\quad e=\frac{n}{2}RT,\quad s=R\ln\left(T^{n/2}v\right),
\end{equation*}
where $R$ is the universal gas constant and $n\ge3$ is the degree of freedom.

The differential quadratic form $\kappa|_{L}$ is of the form
\begin{equation*}
\kappa|_{L}=-\frac{Rn}{2T^{2}}dT\cdot dT-Rv^{-2}dv\cdot dv.
\end{equation*}
It is negative on the entire manifold $L$ and therefore system (\ref{sys}) in case of ideal gases is of hyperbolic type for any process $l\subset L$.

Let us assume that the flow of the gas is adiabatic, i.e. $s(t,x)=s_{0}$. This allows us to express all the thermodynamic variables in terms of $v$ or, equivalently, in terms of $\rho$:
\begin{equation*}
T(\rho)=\exp\left(\frac{2s_{0}}{Rn}\right)\rho^{2/n},\quad p(\rho)=R\exp\left(\frac{2s_{0}}{Rn}\right)\rho^{2/n+1}.
\end{equation*}
Therefore the function $A(\rho)=A_{0}\rho^{m}$, where
\begin{equation*}
A_{0}=\sqrt{R\left(1+\frac{2}{n}\right)\exp\left(\frac{2s_{0}}{Rn}\right)},\quad m=\frac{1}{n}-1.
\end{equation*}
Solution for the velocity has form (\ref{usol1}) and for the density (\ref{rhosol}) is written as
\begin{equation*}
\begin{split}
&\frac{A_{0}^{2}\rho^{2m+1}(2m\rho+2m\alpha_{3}+\rho+2\alpha_{3})}{2\alpha_{2}(m+1)(2m+1)}+\frac{(\alpha_{1}\alpha_{3}-\alpha_{0})^{2}}{2\alpha_{2}(\rho+\alpha_{3})^{2}}+{}\\&+\frac{2x(\rho+\alpha_{3})^{2}-2t\left(\rho^{2}\alpha_{1}+\alpha_{3}(\alpha_{0}+2\rho\alpha_{1})\right)-\rho\alpha_{2}(\rho+2\alpha_{3})t^{2}}{2(\rho+\alpha_{3})^{2}}=0.
\end{split}
\end{equation*}
Equations (\ref{caus}) for the caustic in case of ideal gas and adiabatic process have the form
\begin{eqnarray*}
x(\rho)&=&-\frac{A_{0}^{2}}{\alpha_{2}}\left(\frac{\rho^{2m+2}}{2m+2}+\frac{\rho^{2m+1}}{2m+1}\right)+\frac{\rho(\rho+2\alpha_{3})(\rho+\alpha_{3})^{2}A_{0}^{2}\rho^{2m}-\alpha_{3}^{2}\alpha_{1}^{2}+\alpha_{0}^{2}\pm2\alpha_{0}(\rho+\alpha_{3})^{2}A_{0}\rho^{m}}{2\alpha_{3}^{2}\alpha_{2}},\\
t(\rho)&=&\frac{\pm A_{0}\rho^{m}(\rho+\alpha_{3})^{2}-\alpha_{1}\alpha_{3}+\alpha_{0}}{\alpha_{3}\alpha_{2}}.
\end{eqnarray*}
Finally, the potential $H(\rho,t)$ for the case of ideal gas is
\begin{equation*}
\begin{split}
H(\rho,t)&=\frac{(\alpha_{2}\rho t-\alpha_{1}\alpha_{3}+\alpha_{0})\left(\alpha_{1}\alpha_{3}^{2}+\alpha_{3}((t\alpha_{2}+2\alpha_{1})\rho-\alpha_{0})-2\rho\alpha_{0}\right)}{2\alpha_{2}(\rho+\alpha_{3})^{2}}-{}\\&-\frac{A_{0}^{2}}{\alpha_{2}}\left(\frac{\rho^{2m+3}}{2m+3}+\frac{\alpha_{3}\rho^{2m+2}}{2m+2}\right).
\end{split}
\end{equation*}

\begin{figure}[h]

%\sidecaption
\begin{minipage}[h]{0.32\linewidth}
\center{\includegraphics[scale=0.28]{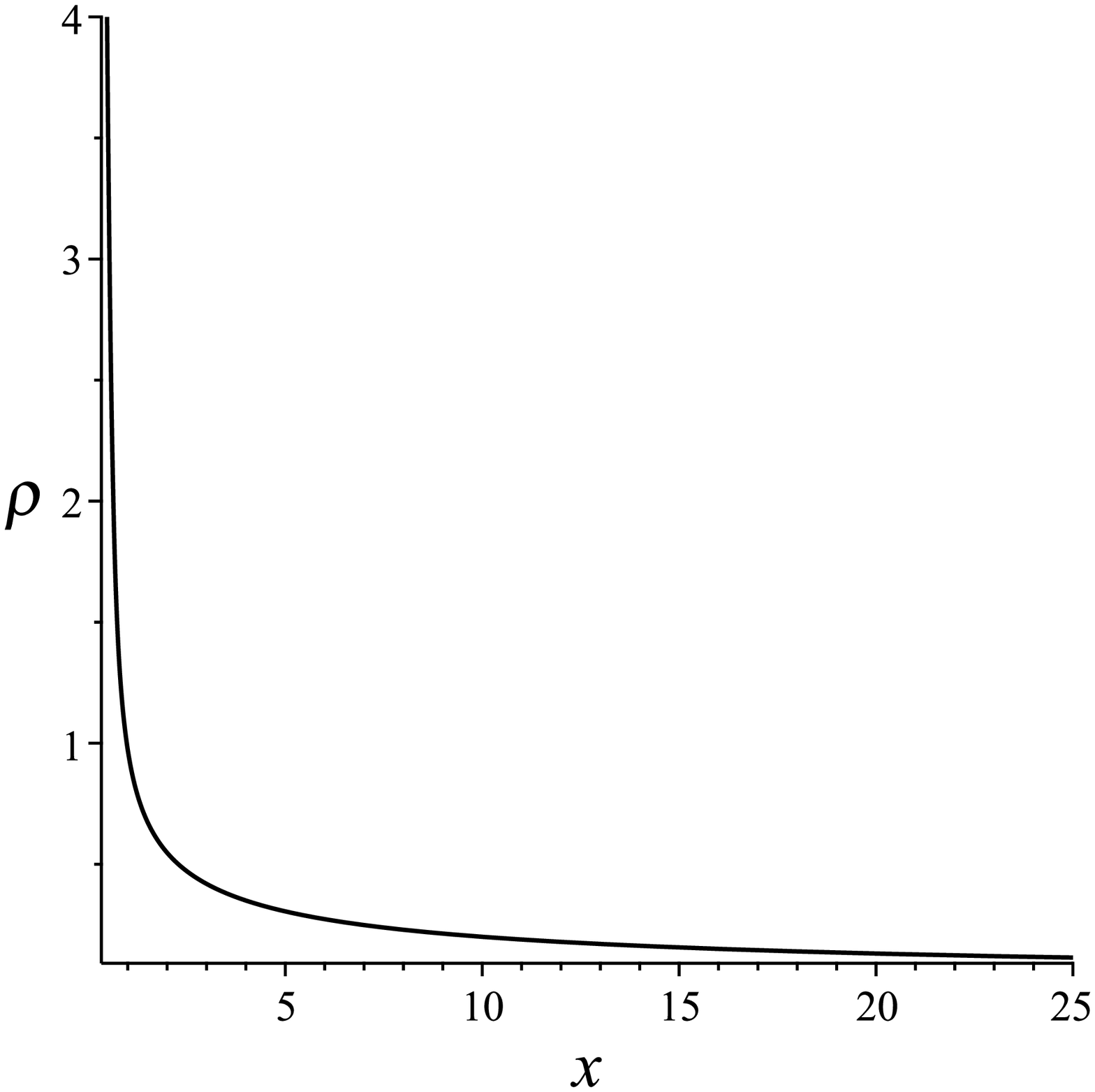}}
\end{minipage}
\hfill
\begin{minipage}[h]{0.32\linewidth}
\center{\includegraphics[scale=0.28]{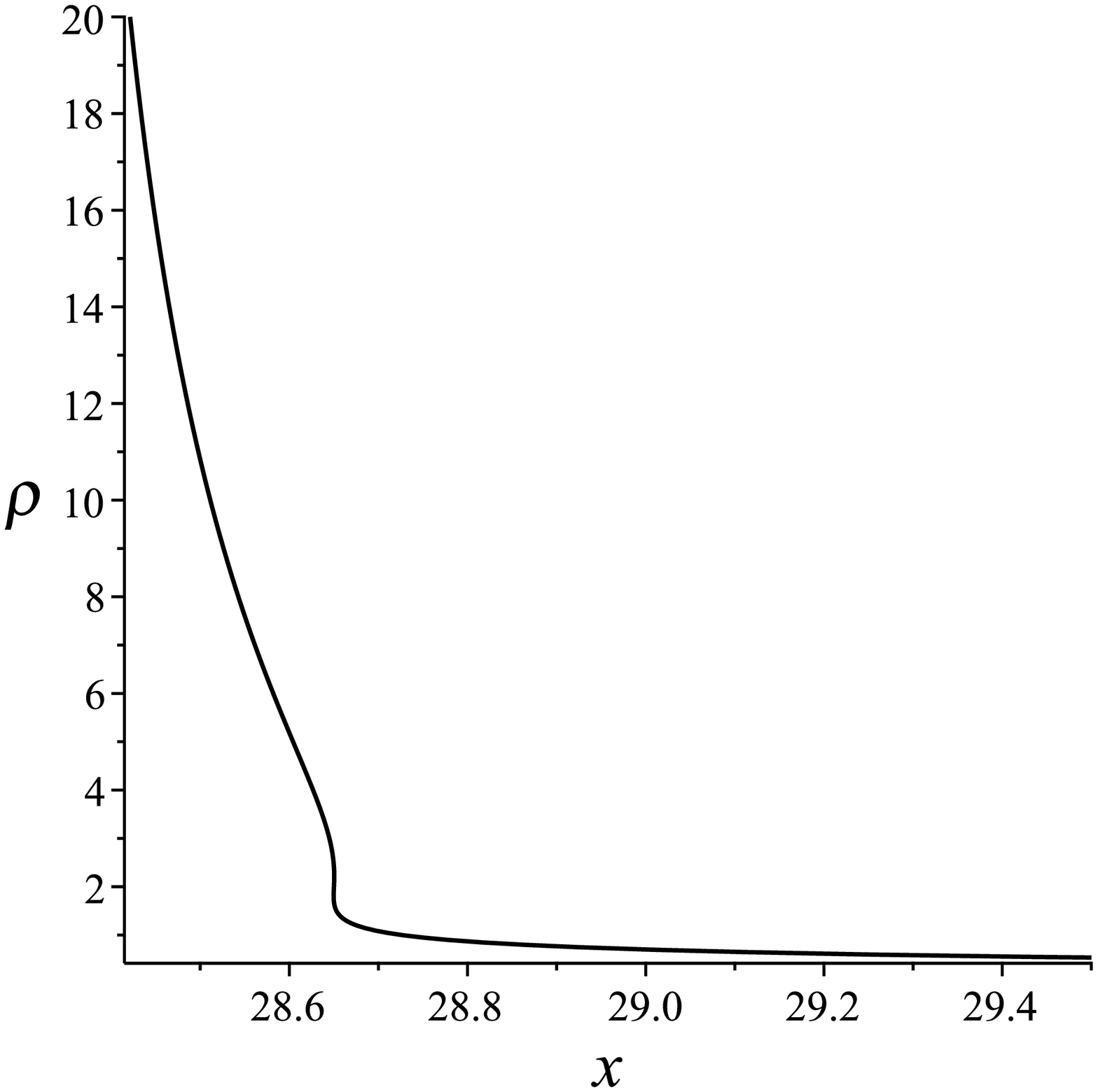}}
\end{minipage}
\hfill
\begin{minipage}[h]{0.32\linewidth}
\center{\includegraphics[scale=0.28]{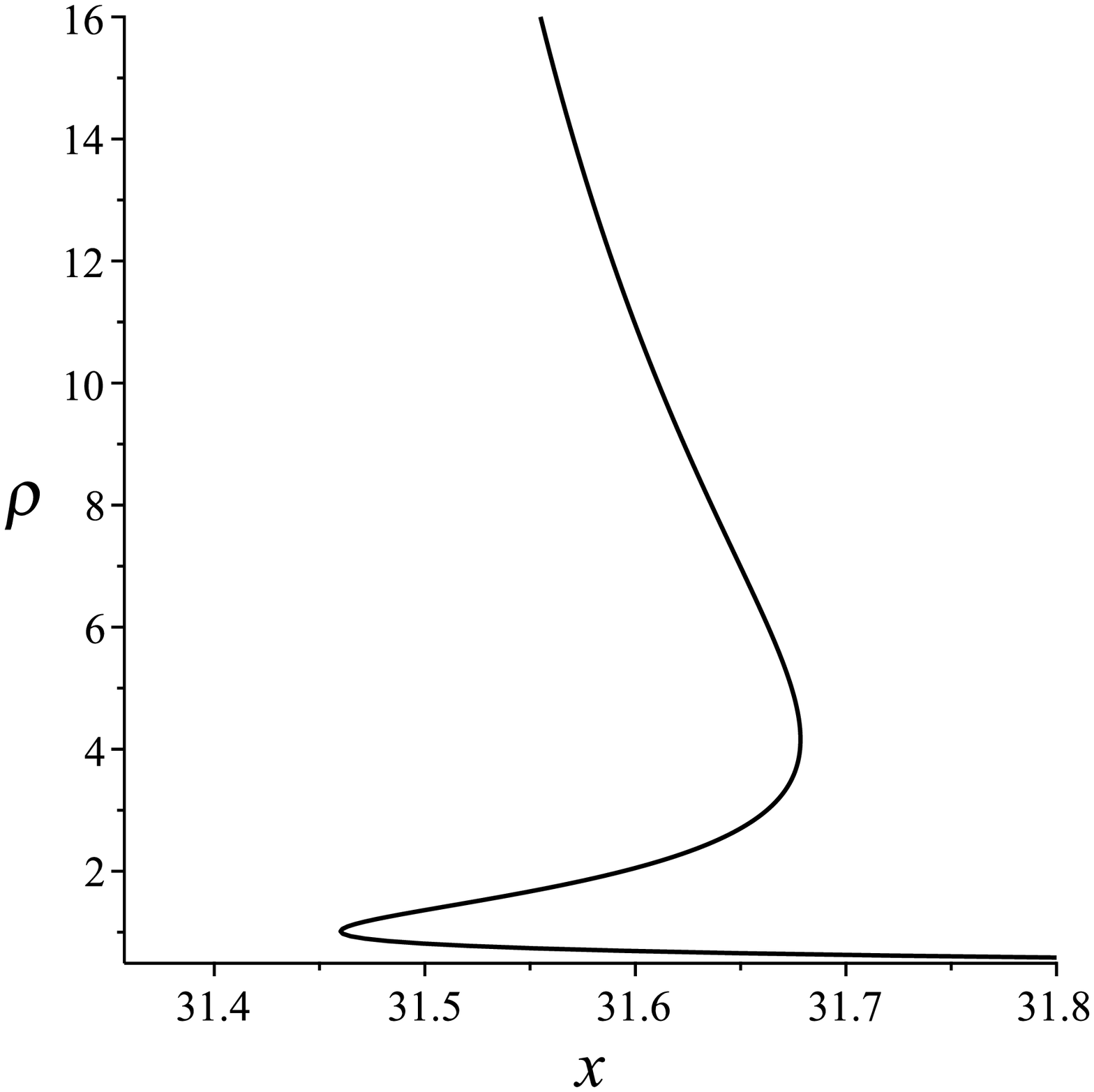}}
\end{minipage}
\caption{Graph of the density in case of $n=3$ for time moments $t=0$, $t=6.58$, $t=7$}
\label{mult}
\end{figure}

Sections of the multivalued solution for various time moments are shown in figure~\ref{mult}.

\begin{figure}
\centering
\includegraphics[scale=.4]{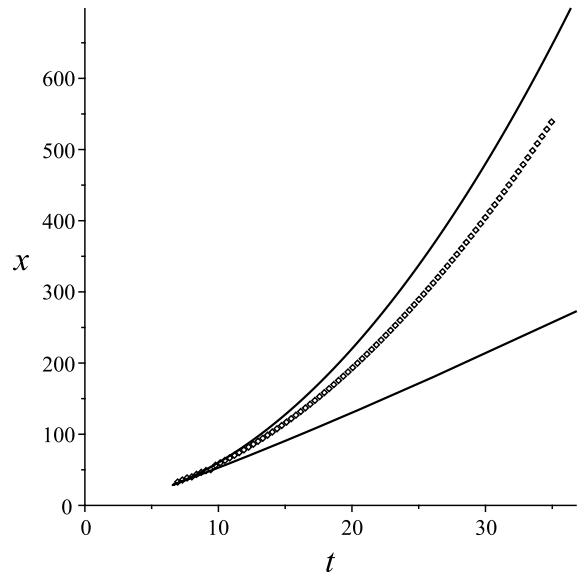}\\
\caption{Caustic (line) and shock wave front (points) for $n=3$}
\label{shock}
\end{figure}

Caustic and shock wave front in the plane $(t,x)$ are shown in figure~\ref{shock}.

\begin{acknowledgments}
This work was partially supported by the Russian Foundation for Basic Research (grant 18-29-10013).
\end{acknowledgments}

% Text of article ends here.
% The Bibliography
%

\end{document}